\documentclass[12pt]{iopart}

\usepackage{iopams}  
\usepackage{url}
\usepackage{graphicx}
\usepackage[utf8]{inputenc}
\DeclareUnicodeCharacter{20AC}{\EUR}
\usepackage{eurosym}
\usepackage{caption}

\begin{document}

\title[For an early introduction of quantum mechanics concepts]{For an early introduction of quantum mechanics concepts in physics curriculum}

\author{Giovanni Organtini}

\address{Sapienza Universit\`a di Roma \& INFN-Sez. di Roma\\Piazzale Aldo Moro 5 - 00185 ROMA (Italy)}
\ead{giovanni.organtini@uniroma1.it}
\vspace{10pt}
\begin{indented}
\item[]July 2018
\end{indented}

\begin{abstract}
In this paper we suggest to anticipate the introduction of concepts such as quantum state and the operators connected to their transformations well in advance of what is usually done. 
\end{abstract}

\vspace{2pc}

{\em Submitted to: }Phys. Ed.
%

\section{Introduction}
With the advent of special relativity~\cite{einstein}, physics knew one of the deepest revolutions in its history. The revolution we are talking about does not consist in the fact that concepts like space and time, till then considered absolute, cannot be taken as such. In our opinion, the most important revolution happening with the formulation of special relativity consists in the fact that {\em reality} cannot be considered independent from {\em measurements}. In this respect, the revolutionary breadth of special relativity is far more important than that of quantum mechanics, although this is little recognised.

Special relativity is based on the experimental observation according to which the speed of light appears to be the same in every reference frame. This observation, to all intents and purposes, is incomprehensible,  in the sense that it is completely extraneous to common sense. In order to accept the conclusions of special relativity, one must accept the fact that Galilean transformations are wrong and that, irrespective of the fact that it appears meaningful or not, what we measure is what we should consider as {\em real}. 

A frequent question posed by students when we teach them about length contraction and time dilation is: are those {\em real} effects or do they just {\em appear} as such? In other words, students are still considering the possibility that time and space are in fact absolute, however, when we measure them, they {\em appear} to be relative, possibly because of the technologies used.

Any physicist considers such a question meaningless: time and space are in fact relative to the observer. The reason for considering such a question meaningless is that, even if time was absolute, there is no way to make an experiment to prove that. Hence, what we measure must be the reality. A reality other than the experimental one cannot exist for physics.

The same reasoning applies to quantum mechanics (even if one of its own creators, Albert Einstein, refused to push his own arguments, adopted in the case of the formulation of the relativity theory, to its very consequences~\cite{einstein-dices}). Despite it may appear meaningless and contrary to common sense, quantum mechanics can predict the results of any experiment with an incredible level of accuracy. Then, irrespective of what we consider reasonable, the Universe works as quantum mechanics predicts. As a consequence, quantum mechanics correctly describes reality, despite someone, especially in the past, considered it 
incomplete, if not completely wrong.

The revolution of modern physics then consists in considering the reality coincident with the set of measurements, while in classical physics reality is somewhat independent on the experimental grounds of physics. This is, in our opinion, the most important difference between classical and modern physics (including special relativity and quantum mechanics in the latter); it is the habit to classical physics that makes the latter less acceptable. 

We argue that changing the way in which we teach classical physics, the transition to modern physics can be much smoother. In fact, some of the most controversial concepts in quantum physics, can be greatly anticipated to the very early stages of learning physics, helping in accepting the conclusions of quantum physics in a less traumatic way.

\section{The concept of the state}
A very common statement in classical physics is that the state of a pointlike particle is given when its position $\vec{x}$ and its velocity $\vec{v}$ are known. 

On the other hand, we could not find a single textbook in which is explained what is intended for {\em state}.

Our proposal is to define the state of a physical system starting from the definition for the word ``state'' given in the vocabulary~\cite{state}, according to which a {\em state} is a {\em mode or condition of being}. In other words it describes, as precisely as possible, which are the conditions in which a given system can be observed at a given time.

In order to provide such information for a pointlike particle, it is enough to provide just few numbers: one for its mass, three for its position and three for its velocity. There is no other meaningful quantities to express the {\em mode or condition of being} of such a particle, in the sense that any other quantity that can be measured or predicted for such a particle, either is meaningless or can be derived from the knowledge of the latter. 

For example, the particle momentum $\vec{p}=m\vec{v}$ can be computed directly from the state variables. If it is taken as a state variable, either $m$ or $\vec{v}$ can be removed from it, being superfluous. The particle acceleration can be predicted from the knowledge of the forces acting on it (that are not part of the system we are describing with the particle). The same applies to its angular momentum with respect to a given (external) point.

Its intrinsic angular momentum, on the other hand, is something that is meaningless for a pointlike particle, as well as its color (pointlike particles do not have a surface), its temperature or its electric resistance.

Once the state of the particle is known, the state of such a particle at a different time can be inferred from what we call ``physics laws''. 

In short, the state of a system can be defined as the set of independent measurements that must be provided to predict, known the boundary conditions, the state of that same system at a different time. Physics laws are equations allowing the computation of the state of a system at a different time with respect to the one at which its state is known.

Manifestly, such a discussion can be made as early as in the first course of mechanics. 

It is worth repeating the discussion at various stages of a physics course. For example, another case in which the word {\em state} frequently appears in physics textbooks is in the study of the laws of ideal gases, where the ideal gas law is often called the gas equation of {\em state}$\,$:

\begin{equation}
pV = nRT\,.
\end{equation}
In this case the state is implicitly given when three of the four quantities (pressure $p$, volume $V$, quantity of gas $n$ and temperature $T$) appearing in the equation are known.

Again, each of the state variables can be predicted once the initial state is known and external conditions applies. For example, if forces act on the container reducing, e.g., its volume, the pressure and the temperature of the gas change accordingly depending on external conditions. 

It is worth noting that, as long as we speak about a gas and we do not interpret it as a set of pointlike particles using the relatively recent kinetic theory, the state of the system is not given in terms of positions and velocities, but, in contrast, in terms of pressure, volume and temperature.

For a gas, terms like position and velocities are meaningless, since there is no way to measure the position of a gas, nor its velocity. A gas is contained in a volume: it does not hold a position. Since a gas, when at the equilibrium, can just sit {\em still} inside the volume, it is impossible to define a procedure to measure its velocity. Even if the gas expands irreversibly as in the Joule experiment, there are no procedures to provide a measurement of its {\em speed}. In fact there is not even a definition of it. As a consequence, the state of a gas cannot be given in terms of position and velocity, but only in terms of quantities that can be measured on it independently: pressure, volume and temperature, for example.

Similar considerations apply in other fields, too. Just to make one example in an unusual field, while introducing the electric current, one can provide the state of a wire whose ends are connected to the plates of a charged capacitor providing the electric current $I_0$ flowing in the wire at $t=0$, its resistance $R$ and the capacitance $C$ of the capacitor. Once these quantities are known, the state of the wire can be told at any time. The capacitance $C$ can be taken as a parameter governing the way in which the state changes, just as a force determines how the state of a pointlike particle changes or the contact with en energy reservoir at fixed temperature determines how the state of a gas transforms. Hence, the state of a wire can be given in terms of $I_0$ and $R$. In other terms, the Ohm's Law is a sort of equation of state for a conductor.

A final remark about the state: the {\em state of being} of, e.g., a wire, may include the colour of its sheathing. Even for a gas, the color can be part of its description. The reason for which we usually do not include it in the state is that it is not relevant for the transformations we are interested in. In order to describe a chemical reaction, however, in which a gas in a container changes its color, the latter can be interesting and must be included in it.

\section{Introducing the operators}
Moreover, one can also introduce as early as possible the concept of an operator as a formal way to describe the effect of something external to the system under investigation which causes a change of its state. An external force $\vec{F}$, for example, transforms a state that can be represented as $\left|\vec{x}, \vec{v}\right\rangle$ at time $t=0$ into a state 

\begin{equation}
\left|\vec{x}+\vec{v}t+\frac{\vec{F}}{2m}t^2,\,\vec{v} + \frac{\vec{F}}{m}t\right\rangle = {\cal O}\left|\vec{x}, \vec{v}\right\rangle
\end{equation}
at time $t$. The transformation from the state at time $t=0$ and the one at time $t\ne 0$ is represented by the application of an operator ${\cal O}$ that provides a transformation of the variables in the state according to some rule that can be inferred from observations, expressed in terms of physics laws. 

Putting a gas, whose state is given by $\left|p, V, T\right\rangle$, contained in a rigid container in contact with a temperature reservoir $T_r$ causes its state to change into a new one

\begin{equation}
\left|p'=\frac{nRT_r}{V}, V, T_r\right\rangle = {\cal O}\left|p, V, T\right\rangle\,.
\end{equation}
In any case the physics law can be regarded as the application of an operator that associates a state (i.e. an element of a set) to another state (i.e. another element of the same set). Often, classical physics operators can be represented by a scalar or vectorial function acting on the space of the state variables. However, that may not be always the case.

This is the case as long as we pretend that physical quantities are real numbers that can attain any value, and as long as we believe that physical quantities are continuous and deterministic functions. 

\section{The concept of a force}
The intuitive concept of a force is something that either pull or push. In general such a model is appropriate for pointlike particles because our eyes, in fact, ``measure'' distances. Forces on pointlike particles (or bodies that can be modelled as pointlike particles) just move their point of application when they don't cancel each other. In other words, these kind of forces alter the state of pointlike particle. However, in Newtonian dynamics, forces are responsible for producing accelerations, i.e. for modifying the state of motion of bodies. Then, the observation of a displacement of a pointlike particle cannot be ascribed to forces, unless a change is its velocity is observed. The intuitive concept of a force is what causes the observed problems in conceptual understanding of Newtonian dynamics~\cite{FCI}. The old Aristotelian concept, according to which a moving particle is always subject to a force, still persists in many students.

The application of a force on a gas is only apparently similar to what happens to pointlike particles. While the force can still be regarded as something that push or pull, it cannot be represented by a vector. At least, it cannot be always represented by just one vector. Forces like those applied to a pointlike particle can be applied to the gas container. The container, in turn, exerts a force on the gas. However, while the force on the container can be applied on just one of its point, the forces exerted on the gas by the container must be applied to something that has a volume and is always directed perpendicularly to the container's walls. Moreover, the effect of forces on the container is distributed over its whole surface. In this case there is no acceleration, nor a displacement of the gas (there is, of course, a displacement of the particles composing the gas that, however, are not the gas as a whole). That is why one is forced to introduce the concept of pressure, defined as a scalar quantity that can be measured combining the measurement of a force and of a surface.

A similar problem appears when we talk about electric currents in wires. Until we do not have a model of electricity in terms of particles subject to the electromotive force, one cannot clearly identify any force acting on the wire. However, it is clear that some kind of force must develop when connecting the wire to a capacitor. In fact, the temperature of the wire, i.e. its internal energy, increases and the first principle of thermodynamics states that the increment can only be due to the work done by the capacitor. From the definition of work it follows that the capacitor must apply some sort of force to the wire, that, however, does not move at all nor changes its shape or size. 

From the above observations it follows that forces are those entities that tend to cause a change in the state of a system. Forces are not necessarily vectors. They are vectors only when they result in pulling or pushing a point. In other cases they must be described in some other way.

The operators mentioned above, in fact, are a possible way to express the effect of a force. One can clearly see that, in the above examples, vectors only appear when describing the effect on the state of pointlike particles, just because the state is itself represented as a pair of vectors.

This definition of force allows us to easily overcome a very common problem that occurs when citing the four fundamental forces of Nature. While gravity, electromagnetism and the strong force can be easily described in terms of something that push or pull, the weak force, causing certain radioactive nuclear decays, cannot be described in these terms. Students have great difficulties in understanding why we need a force to model beta decay, for example. In such phenomena there is nothing that accelerates, nor something that moves. The emitted particles, of course, do that, but the force we are talking about is not those causing the motion of the emitted particles, but those that make the decay to happen.

Adopting our point of view, the need of a force becomes clear: at a given time we measure the state of a system being composed, let's say, of just an atom, whose atomic number and atomic mass are, respectively, $Z$ and~$A$, i.e., $\left|Z,A\right\rangle$. At a later time, the state of the system has changed into a new state

\begin{equation}
\left|Z\pm 1, A, e, \nu\right\rangle = {\cal O}\left|Z,A\right\rangle\,.
\end{equation}
A change in the state requires the intervention of a force (or, better, an interaction) represented by the action of the operator ${\cal O}$ on the initial state.

An even better introduction to this problem consists in making a difference between {\em interactions} and {\em forces}. Interactions are produced by a source (the mass for gravity, the electric charge for electricity, the electric current for magnetism, the color for the strong interaction and the hypercharge for the weak interaction) and, depending on the way in which the state is described, may results in the appearing of a force. This way, forces defined as ``fictitious'', like those appearing in non--inertial reference frames, appear to be as real as non--fictitious ones. In fact, this description makes it possible to overcome another difficulty in understanding Newtonian dynamics, coming from our experience that, as a matter of fact, fictitious forces are among the few forces that we really {\em feel}\footnote{Being our senses immersed in the gravitational field, gravity is not usually perceived as a force, in fact.}.

The operator ${\cal O}$ must be able to transform the initial state from $\left|Z,A\right\rangle$ to $\left|Z\pm 1, A\right\rangle$, but it also must {\em create} a pair of new particles (the electron $e$ and the corresponding neutrino~$\nu$). Forces, then, can do much more than just pushing or pulling. It is worth noting that, despite this possibility was implicitly contained already in the early quantum mechanics, until the work of Enrico Fermi~\cite{fermi} the whole scientific community believed that the electrons emitted in beta decay pre-existed in the nuclei. That was causing many problems to the interpretation of the phenomenon. The adoption of this point of view, in contrast, was proven to be correct in the sense that almost all of the predictions made using the Fermi's theory turned out to be correct. The possibility to {\em create} particles not existing in the initial state was considered shocking at the beginning, but as a matter of fact there is nothing in classical physics preventing it, provided that forces (interactions) are correctly introduced.

\section{From classical physics to quantum mechanics}
The introduction of the weak interactions is just the first step into a gradual formulation of quantum mechanics. Presenting quantum mechanics as an abrupt revolution with respect to classical physics may be fascinating for many reasons. However, it leaves the conviction in students that there must be something wrong in it, being so much different to what was learnt so far. This was, in fact, the same mood in the scientific community at the very beginning of the new mechanics. With time, we realised that, in fact, quantum mechanics is indeed much more solid than classical physics. There is no field in physics in which the predictions of the theory are verified at precision levels comparable with those attained in quantum mechanics. A smoother transition between classical physics and quantum mechanics, then, is desirable because it may help in reducing the difficulties that students have in accepting its results.

In fact, we believe that it is even possible to do the opposite, i.e., starting from quantum mechanics to land into classical physics. However, this is a long term programme that we will pursue in the future.

Once the picture described in the previous sections has been consolidated, one can introduce quantum mechanics in many ways. For example, starting from the uncertainty principle, one can see that there is not a solid procedure to measure the position and the velocity of an electron bound in an atom. Hence, its state cannot be described in these terms. As in the case of the gases and of the wires, we are forced to find other physical quantities that can be used to describe the state of such electrons, i.e. their mode of being.

One can see, for example, that the binding energy of the electron can be obtained by means of the photoelectric effect and it can be experimentally shown that the electrons in all the atoms of each chemical species share the same set of energies. At the same time, one can show that the angular momentum of the electrons are quantised and that the value of such a quantity is independent on the energy. As a consequence, not differently from classical physics, the state of electrons in atoms must be described by the only meaningful quantities describing their mode of being: $\left|E, J, J_z\right\rangle$ where $E$ is their energy and $J$ and $J_z$ their total angular momentum and their third component, respectively.

The interaction with a field causes the appearance of a force that in turn induces, as usual, a change in the state. Photoelectric effect, Compton scattering and other phenomena can all be described in terms of an interaction causing a change in the state of the electron.

Moreover, assuming this picture, one can easily explain phenomena considered very unnatural, like electron diffraction by two slits. The latter is simply explained by the fact that what emerges from the screen with the slits is not a flux of pointlike particles (we cannot tell the position of each single electron), but something whose state include a given momentum. Such an object is characterised by the fact that at $t=0$ (the time of the crossing), the distribution of the number of electrons is known to be a pair of lines, whose position, however, is not even defined. The evolution of such a state naturally gives rise to the diffraction pattern, not much differently of what happens to a gas expanding against the vacuum in a container. We cannot measure the position and the velocity of the gas because it does not make sense to do that, but we can be sure that the gas flows in the container from the faucet when it is open. The only difference is that, in classical physics, one supposes (with any experimental support) to be allowed to perform a measurement that tells us the position of each single particle in the gas, while in the case of electrons that is not possible.

\section{Testing the hypothesis}
The hypothesis under the above considerations is that quantum mechanics appears to be much more acceptable to students, if they have been prepared to deal with it while learning classical physics. We performed a first test of this hypothesis teaching physics to a class of students in biotechnology, applying the methods outlined above.

The physics course is given during the first semester of the first year, so the students attending the class have just exited the high school. Most of them (84~\%) already received a first introduction to physics, often including elementary quantum mechanics. It is worth noting that the population can be regarded to be almost homogeneous to that of high school students, since, besides the age, those students share with them a very soft interest for physics.

At the end of our course we administered a questionnaire that was answered by 108 students, among which 17 were not exposed to physics in high school. To each question we proposed five possible answers: from ``not at all'' to ``definitely'', with ``no'', ``neutral'' and ``yes'' as possible intermediate values. Table~\ref{tab:is-interesting} shows the answers to the question ``do you believe you understood physics more deeply with the approach adopted in this course''? 35~\% of the respondents answered positively, 45~\% are neutral, while 20~\% did not agree. 

To the question ``Do you think that the way the course is designed is more difficult than a traditional course?'' those who answered positively were~17~\%, while 40~\% are neutral and 43~\% did not agree.

Interestingly enough, those who declared that physics were not among the disciplines they liked while at school, changed from~42~\% to~18~\%. Such a result depends on many variables, however, since the whole course was designed around the principles outlined in the previous sections, we believe that a great part of this achievement is to be ascribed to their application, who made classical physics more and more appealing as the students understand the connection between classical and modern physics.

We intend to continue investigating on this point, to understand what, in particular, make students change their mind about physics among the various actions taken during the course. 

\section{Conclusion}
We propose to introduce quantum physics terminology and principles rather early in the study of classical physics. In particular, a rather more formal definition of the state of a system and  the introduction of operators as something that produces a change in the state, to be identified with the presence of some interaction, might be made at the very beginning of a physics course and greatly help in introducing quantum mechanics.

A very preliminary investigation has been done about student's perception of the subject. First results are encouraging, hence we look forward to search for partners for a physics education research aiming at identifying how the proposed approach can effectively increase the interest and the comprehension of physics in high school and college students.

\end{document}